\documentclass[pra,twocolumn,amsmath,amssymb,showpacs,superscriptaddress]{revtex4}
\usepackage{braket}
\usepackage{graphicx,epstopdf}

\renewcommand{\Im}{\mathfrak{Im}}
\newcommand{\dd}{\ensuremath{\, \textnormal{d}}} 
\newcommand{\trace}{\operatorname{tr}}
\newcommand{\old}{\operatorname{old}}
\newcommand{\new}{\operatorname{new}}
\newcommand{\Op}[1]{\boldsymbol{\mathsf{\hat{#1}}}}
\newcommand{\unity}{\leavevmode\hbox{\small1\kern-3.3pt\normalsize1}}

\begin{document}

\title{Robustness of high-fidelity Rydberg gates with single-site
  addressability} 

\author{Michael H. Goerz}
\thanks{These authors contributed equally.}
\affiliation{Theoretische Physik, Universit\"{a}t Kassel,
  Heinrich-Plett-Str. 40, D-34132 Kassel, Germany} 

\author{Eli J. Halperin}
\thanks{These authors contributed equally.}
\affiliation{Theoretische Physik, Universit\"{a}t Kassel,
  Heinrich-Plett-Str. 40, D-34132 Kassel, Germany} 
\affiliation{Department of Chemistry, 
  University of California, Berkeley, California 94720, USA}

\author{Jon M. Aytac}
\affiliation{Department of Chemistry, 
  University of California, Berkeley, California 94720, USA}

\author{Christiane P. Koch}
\affiliation{Theoretische Physik, Universit\"{a}t Kassel,
  Heinrich-Plett-Str. 40, D-34132 Kassel, Germany} 

\author{K. Birgitta Whaley}
\affiliation{Department of Chemistry, 
  University of California, Berkeley, California 94720, USA}

\begin{abstract}
  Controlled phase (CPHASE) gates can in principle be realized with trapped
  neutral atoms by making use of the Rydberg blockade.  Achieving the ultra-high
  fidelities required for quantum computation with such Rydberg gates is however
  compromised by experimental inaccuracies in pulse amplitudes and timings, as
  well as by stray fields that cause fluctuations of the Rydberg levels.  We
  report here a comparative study of analytic and numerical pulse sequences for
  the Rydberg CPHASE gate that specifically examines the robustness of the gate
  fidelity with respect to such experimental perturbations.  Analytical pulse
  sequences of both simultaneous and stimulated Raman adiabatic passage (STIRAP)
  are found to be at best moderately robust under these perturbations. In
  contrast, optimal control theory is seen to allow generation of numerical
  pulses that are inherently robust within a predefined tolerance window. The
  resulting numerical pulse shapes display simple modulation patterns and their
  spectra contain only one additional frequency beyond the basic resonant
  Rydberg gate frequencies.
  Pulses of such low complexity should be experimentally feasible, allowing gate
  fidelities 
  of order 99.90 - 99.99\% to be achievable under realistic experimental conditions. 
\end{abstract}

\date{\today}
\maketitle

\section{Introduction}
\label{sec:intro}

Rydberg states of trapped neutral atoms provide an attractive platform for
realizing quantum information processing, offering a strong interaction between
relatively distant and otherwise non-interacting atoms that may be switched on
and off with focused lasers~\cite{SaffmanRMP10}.  Proposals have been made for
quantum gates with both addressable and non-individually addressable single atom
qubits~\cite{JakschPRL00} as well as with atomic
ensembles~\cite{lukin2001dipole,beterov2013quantum}.  These schemes typically
employ resonant excitation and make use of the Rydberg blockade to generate
controlled phase relationships between logical qubit states that are typically
defined as hyperfine states of the ground electronic atomic manifold.  Progress
in trapping and manipulating single atoms in dipole traps and optical tweezers
has enabled experimental validation of the key theoretical concepts of the
Rydberg blockade~\cite{UrbanNatPhys09,GaetanNatPhys2009}, as well as subsequent
use of this to generate entanglement between trapped atoms in these
configurations~\cite{WilkPRL10,IsenhowerPRL10}.   The latter study also
demonstrated a low fidelity version of a controlled not (CNOT) gate based on the
Rydberg blockade.  Parallel to this, several groups have developed the
capability to form arrays of trapped atoms in optical lattices that are
characterized by single site occupancy and
addressability~\cite{bergamini2004holographic,nelson2007imaging,whitlock2009two,bakr2009quantum,kruse2010reconfigurable,weitenberg2011single},
thereby opening the path to large scale quantum information processing with
atomic qubits.  

Despite these conceptual and experimental advances, realization of high fidelity
quantum logic gates between such trapped neural atoms has remained elusive, due
to the significant challenges involved in coherently controlling and
manipulating optically trapped atoms. Two-qubit gates relying on controlled use
of dipolar interactions between atoms in Rydberg states have the potential of
being fast, but are subject to a number of intrinsic and technical sources of
error that can restrict both the achieved fidelity and speed of operation.  An
important source of intrinsic error specific to Rydberg gates is the lifetime of
the atoms in the Rydberg states while technical errors may derive from a number
of experimental factors, as discussed recently in~\cite{zhang2012fidelity}. For
the non-individually addressable implementation of the Rydberg gate protocol
in~\cite{JakschPRL00}, atomic motion can also play a significant role in
limiting the fidelity~\cite{GoerzJPB11,murphy2011towards}.  The role of these
and other factors limiting gate fidelities have been studied theoretically for
Rydberg gate schemes involving both analytic pulse
sequences~\cite{brion2007implementing,zhang2012fidelity} and, for the
non-addressable protocol of ~\cite{JakschPRL00}, numerically optimized
pulses~\cite{GoerzJPB11,murphy2011towards}. These studies indicate that gates
with errors of the order of 10$^{-3}$ might be achieved with suitable choice of
atoms and qubit levels.  However, no study of the {\em robustness} of two-bit
gates with respect to errors has been made, although such robustness with regard
to fluctuations of both intrinsic and technical parameters is a critical
desiderata of experimental studies.  In this work we remedy this with
a systematic study of the robustness of both analytic and numerical pulse
sequences with respect to the primary technical fluctuating parameters, namely
pulse timing, pulse amplitude, and two-photon
detuning.

Another desiderata for quantum information processing is the realization of fast
gates.  While proposals have been made to mitigate the effects of intrinsic
errors in Rydberg gates using adiabatic passage
techniques~\cite{gaubatz1990population}, the resulting pulse sequences typically
result in relatively long gate times of $\mu$s or
longer\cite{moller2008quantum,beterov2013quantum}, which is disadvantageous for
quantum computation schemes that generally require large numbers of gates.
Prospects for achieving gates on ns timescales have been reviewed
in~\cite{MuellerKochSpIssue11} and for the non-addressable protocol
of~\cite{JakschPRL00}, optimal control theory has been used to characterize
bounds on the shortest possible gate time~\cite{CanevaPRL09}, corresponding to
a ``quantum speed limit'' for performing the gate~\cite{GiovannettiPRA03}.

The remainder of the paper is constructed as follows.  Section~\ref{sec:model}
summarizes the atomic level structure and qubit model, as well as basic
components of the CPHASE gate implementation with 
Rydberg states of individually addressable atoms.
For technical reasons, the transition to the Rydberg level is via an
intermediary state.
In Section~\ref{sec:analytic} we first analyze the performance of three forms of
analytic pulse sequences.
The first is the original $\pi - 2\pi-\pi$ sequence
of~\cite{JakschPRL00}, where each pulse consists of a simultaneous pulse pair
realizing a two-photon transition to a Rydberg state.
The second is a fully adiabatic version of this, in which each simultaneous pulse pair is
replaced by a STIRAP pulse pair, and the third is a mixed scheme in which only the $\pi$
pulses are replaced by STIRAP pulse pairs.  These different schemes are then
compared in their robustness with respect to intrinsic experimental parameters.
We find that the mixed scheme is the most robust of these analytic approaches,
due to its selective use of STIRAP on the control qubit only.  However, all
STIRAP based schemes require either large pulse amplitudes or exceedingly long
pulse times.  Section~\ref{sec:oct} demonstrates the benefits offered by
numerical optimal control calculations in generating pulse sequences.
We first determine the optimal pulses  for a given pulse duration using the
Krotov method \cite{Konnov99,PalaoPRA03,ReichKochJCP12}
within a density matrix formulation for the open quantum system dynamics, taking
spontaneous emission into account.  Optimization for pulses robust to
fluctuations in pulse amplitude and Rydberg energies (due, e.g., to
stray fields) is then made over an ensemble of Hamiltonians within an
experimentally relevant tolerance window.  We find that optimal control yields
systematically higher gate fidelities than all analytic approaches, showing
improvement of an order of magnitude to reach gate errors of order 10$^{-4}$
for equivalent gate times.  Most importantly, optimal control can deliver gate
performance that is also extremely robust with respect to experimental
fluctuations, with the gate error staying below or at the order of $10^{-3}$
even for large fluctuations.
Using optimal control we can also significantly shorten the total gate duration,
to $\sim$100~ns, approaching the quantum speed limit for these systems, without
loss in either robustness or fidelity.  The resulting numerical
pulse spectra are surprisingly simple and 
allow the
error threshold for fault tolerant computation to be reached at the price of a small
increase in pulse complexity relative to the analytic sequences.

\section{Model}
\label{sec:model}

\begin{figure}[tb] 
  \begin{center}
    \includegraphics{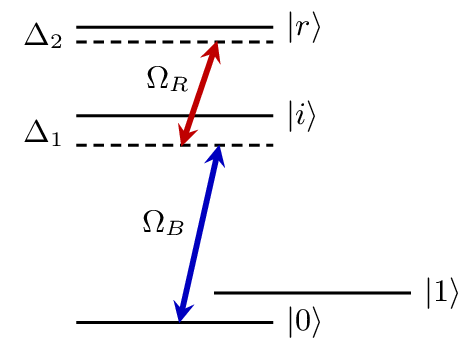}  
  \end{center}
  \caption{(Color online)
    Level scheme for a single atom. The color scheme given here, blue for the
    lower transition and red for the upper one, is used throughout the figures
    of this paper.}
  \label{fig:levels}
\end{figure}
We consider two cesium atoms trapped in an optical lattice with
single-site addressability. 
The qubit states are encoded in hyperfine levels of the ground state, 
$\Ket{0} =\Ket{6\,^2S_{1/2}, F=3}$,
$\Ket{1} =\Ket{6\,^2S_{1/2}, F=4}$. 
For practical reasons, the Rydberg level, here 
$\Ket{r}=\Ket{50D_{3/2}}$, is accessed by a two-photon transition via an
intermediate state, $\Ket{i}=\Ket{7P_{3/2}}$. 
In the basis $\{\Ket{0},\Ket{1},\Ket{i},\Ket{r}\}$,
the Hamiltonian for a single atom, using a two-photon
rotating-wave approximation~\cite{ShoreBook11}, reads
\begin{equation}
  \label{eq:H_1q}
  \Op{H}_\text{1q} = 
  \begin{pmatrix}
      0 & 0 & \Omega_B(t) & 0 \\
      0 & E_1 & 0 & 0 \\
      \Omega_B(t) & 0 & \Delta_1 & \Omega_R(t) \\
      0 & 0 & \Omega_R(t) & \Delta_2
  \end{pmatrix}\,,
\end{equation}
where $\Omega_B (t), \Omega_R (t)$ are the Rabi frequencies of the `blue' and
`red' pulses, cf.\ Fig.~\ref{fig:levels}, and $\Delta_1, \Delta_2$ are the
one-photon and 
two-photon detunings. The two atoms are kept at a distance of
$5 \mu$m such that their interaction is negligible except when both
atoms are in the Rydberg state. 
The Hamiltonian for the two atoms, including their Rydberg
interaction, is written as
\begin{equation}
  \label{eq:2q_ham}
  \Op{H}_{2q} = \Op{H}_{1q} \otimes \openone + \openone \otimes \Op{H}_{1q}
  - u \Ket{rr}\langle rr|\,,
\end{equation}
with interaction energy $u$. The parameters are summarized in
Table~\ref{table:params}. 
\begin{table}
  \begin{tabular}{llcr}
  Single-Photon Detuning & $\Delta_1$ & $=$ & $1.273$ GHz \\
  Two-Photon Detuning    & $\Delta_2$ & $=$ & $0$ MHz \\
  Qubit Energy           & $E_1$      & $=$ & $9.100$ GHz \\
  Interaction Energy     & $u$        & $=$ & $57.26$ MHz
  \end{tabular}
  \caption{System Parameters}
  \label{table:params}
\end{table}
Rabi frequencies of $\Omega_B = 171.5$ MHz
and $\Omega_R = 148.4$ MHz have been implemented for this system and 
values up to $\sim 250$ MHz are expected to be experimentally feasible~\cite{TedRyd} .

Resonant excitation of both atoms to the Rydberg state leads to an
acceleration of the atoms due to the dependence of the Rydberg
interaction strength on interatomic separation~\cite{JakschPRL00}. 
The minimum gate duration 
is then determined either by the inverse of the interaction, $u$, or by
the period of the atomic motion in the trap~\cite{GoerzJPB11}.
The gate duration may be limited further by the inverse of the 
experimentally realizable Rabi frequencies. 

We consider the Rydberg blockade regime which avoids resonant
excitation into $\Ket{rr}$. It corresponds to 
\begin{equation}
  \label{eq:blockade}
  u\gg\Omega_j\quad (j=B,R)
\end{equation}
The original proposal of the Rydberg gate~\cite{JakschPRL00} in this regime
requires the atoms to be individually addressable, and employs
a sequence of three pulses: a $\pi$-pulse on the left atom, resulting
in complete population transfer from $\Ket{0}$ to $\Ket{r}$, 
followed by a $2\pi$-pulse on the right atom and another $\pi$-pulse
on the left atom. If the qubits are initially in $\Ket{00}$, a
non-local phase is accumulated during the middle 
pulse because of the detuning of level $\Ket{rr}$ due to the
interaction, $u$, and we thus can execute a CPHASE gate. 

\section{Analytic pulse sequences}
\label{sec:analytic}

When a resonant two-photon transition is employed via an
intermediate level , the two-level system
$\{\Ket{0},\Ket{r}\}$ for one atom in the original
proposal~\cite{JakschPRL00} is 
replaced by $\{\Ket{0},\Ket{i},\Ket{r}\}$. The $\pi$ and $2\pi$ population
flips can then be realized either with
two simultaneous pulses: where $\Omega_B$ connecting
$\Ket{0}$ and $\Ket{i}$ and $\Omega_R$ connecting $\Ket{i}$
and $\Ket{r}$ are driven contemporaneously; or via a STIRAP process: where $\Omega_R$
acts as a ``Stokes'' pulse, preceding but overlapping $\Omega_B$, the ``pump''
pulse. Both methods may be combined in a mixed scheme, where a STIRAP sequence
is used for the $\pi$ flip acting on the left atom, while the $2\pi$ flip on the right
atom is realized using simultaneous pulses.
The following sections discuss the merits and drawbacks of all three approaches, and
numerically analyze the robustness with respect to pulse timing, fluctuations of
the Rydberg level, and fluctuations of the pulse amplitude.

\subsection{Sequence of three simultaneous pulse pairs}
\label{subsec:JZ}

We first consider the realization of all population transfers using simultaneous
pulse pairs. The pulses are of Blackman shape,
\begin{equation}
  \label{eq:blackman}
  S(t) = \frac{E_0}{2}\left(1 - a - \cos\left(2\pi t/T\right)
  + a\cos\left(4\pi t/T\right)\right)\,, 
\end{equation}
with $a = 0.16$ and $E_0$ the peak amplitude. This pulse shape is essentially
identical to a Gaussian centered at $T/2$ with a width of $\sigma = T/6$, but,
unlike the Gaussian, is exactly zero at $t=0$ and $t=T$. Other pulse shapes are
possible.

\begin{figure}[tb] 
  \begin{center}
    \includegraphics{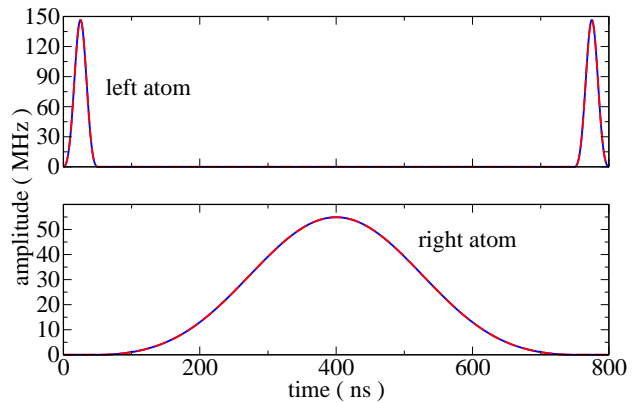}
  \end{center}
  \caption{(Color online)
    Three sequential Blackman pulse pairs implementing a CPHASE gate.}
  \label{fig:jz_pulses}
\end{figure}
A pulse sequence that realizes the two $\pi$-flips on the left atom and one
$2\pi$-flip on the right atom is shown in Fig.~\ref{fig:jz_pulses}. Due to the
large single photon detuning of $1.3$~GHz, the intermediate level can be
adiabatically eliminated. This places a restriction on the pulse amplitude,
\begin{equation}
    \Omega_j \ll \Delta_1 \quad (j=B,R)\,.
\end{equation}
The $2\pi$ pulse is more stringently restricted by the blockade condition in
Eq.~(\ref{eq:blockade}). With the pulse duration being inversely proportional to
the pulse amplitude, both effects result in a quantum speed limit.

\begin{figure}[tb] 
  \begin{center}
    \includegraphics{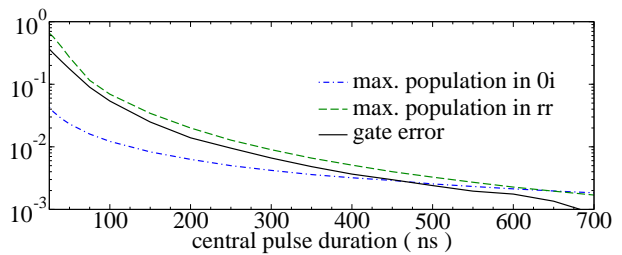}    
  \end{center}
  \caption{(Color online)
    Quantum speed limit for the Rydberg gate using simultaneous
    Blackman pulse pairs. The time window is only that of the center
    $2 \pi$ pulse in the scheme.
    As a measure of the breakdown of the Rydberg blockade,
    the maximum population in the $\Ket{rr}$ state during that pulse is shown,
    as well as the maximum population in the $\Ket{01}$ state, as a measure of
    the breakdown of the adiabatic elimination of the intermediate level.
    Finally, we show the total gate error obtained when combining the center
    $2\pi$ pulse of the given duration with two $50$~ns $\pi$ pulses on the left
    atom.
  }
  \label{fig:QSL}
\end{figure}
Quantitatively, the limitations are illustrated in
Fig.~\ref{fig:QSL} which shows the gate error (black solid line) vs.\ duration
of the middle $2\pi$ pulse, using 
a duration of $50$~ns for the initial and final $\pi$ pulse. The gate error is
defined as $1-F$ where $F$ is the gate fidelity,
\begin{equation}
  F = \frac{1}{20} 
      \left(
        \Big\vert \trace \left[ \Op{O}^\dagger \Op{U} \right] \Big\vert^2
        + \trace \left[ \Op{U} \Op{U}^\dagger \right]
      \right)\,,
 \label{eq:gate_fid}
\end{equation}
with $\Op{O}$ the target CPHASE gate, and $\Op{U}$ the projection of the
time evolution operator onto the logical subspace ($\Op{U}$ is unitary if and only if
there is no loss from that subspace at final time $T$).
The breakdown of adiabatic elimination becomes apparent in the peak
population of the $\Ket{0i}$ state (green dashed line), whereas a
breaking of the Rydberg blockade is observed in 
the peak population in the $\Ket{rr}$ state (blue dot-dashed line). 
Gate errors below $10^{-3}$ are only reached
for pulse durations of $\geq 800$~ns. 
The gate time is dominated by the central $2\pi$ pulse, which must be
sufficiently weak to not break the Rydberg blockade. Already, the pulse
amplitude is remarkably close to the interaction energy, pushing the limits of
condition (\ref{eq:blockade}).
Note that the choice of identical
peak Rabi frequencies for the red and blue laser,
$\Omega_{B,\max}=\Omega_{R,\max}$, is the only ratio
possible to guarantee complete population inversion in a three-level
system using simultaneous pulses when the intermediate level is adiabatically
eliminated ~\cite{ShoreBook11}.

\begin{figure*}[tb] 
  \begin{center}
    \includegraphics{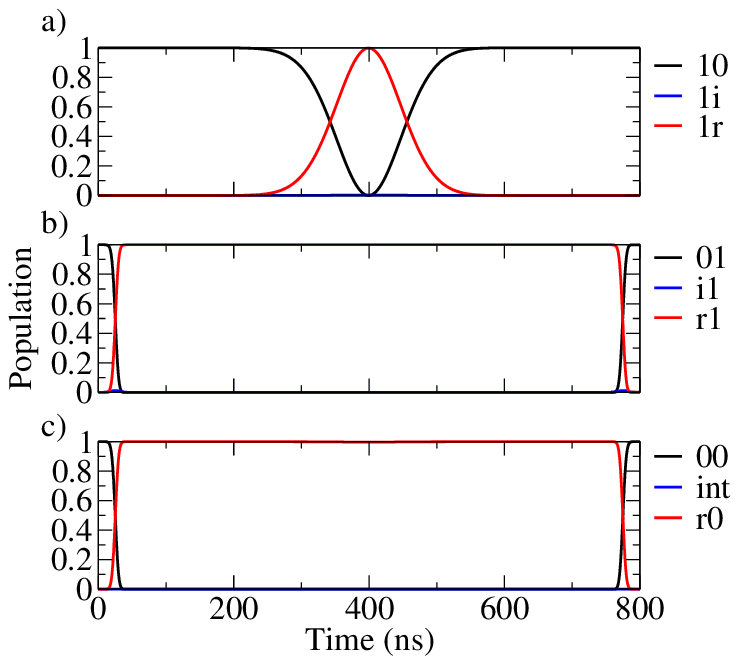}
    \includegraphics{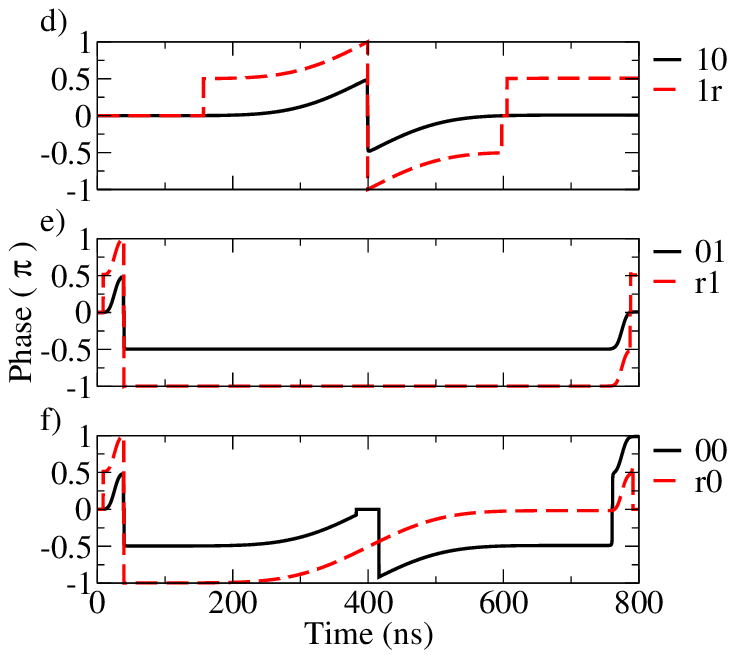}
  \end{center}
  \caption{Population and phase dynamics using the simultaneous pulses
    shown in Fig.~\ref{fig:jz_pulses}}
  \label{fig:jz_dynam}
\end{figure*}
Population and phase dynamics obtained with simultaneous red and blue 
pulses is shown in figure~\ref{fig:jz_dynam}. As described in
section~\ref{sec:model}, the population undergoes a $\pi$ Rabi cycle
on the left atom, followed by a $2\pi$ pulse 
on the right atom, followed by a $\pi$ pulse on the left atom,
cf. Fig.~\ref{fig:jz_dynam} (a,b,c). The intermediate level receives 
almost no population and thus,
for this time scale, spontaneous decay is not an issue.  As can be
seen from Fig.~\ref{fig:jz_dynam} (f), 
the non-local phase is accumulated in the $\Ket{00}$ state entirely
during the central $2\pi$ pulse. Although the Rydberg blockade
is not broken, and the population remains 
in $\Ket{r0}$, the state accumulates an additional phase due to the
detuned pulse driving the transition out of $\Ket{r0}$. This additional
phase is critical for the success of the gate. 

\subsection{Sequence of STIRAP pulse pairs}
\label{subsec:stirap}

STIRAP is a popular scheme to achieve population transfer in
three-level systems, avoiding population in the intermediate level at all times
\cite{BergmannRMP98}.
It is based on adiabatically following a
dynamic dark state that does not contain an $\Ket{i}$-component.
In our setup, the scheme for transferring population from $\Ket{0}$ to
$\Ket{r}$ is realized by first switching on the red laser, acting as
a ``Stokes'' pulse, followed by the blue laser, acting as the ``pump'' pulse.
The two pulses must overlap, but the process is robust with respect to the
laser amplitude and the exact overlap of the pulses, as long as the
condition for adiabatic following, roughly given
by~\cite{BergmannRMP98}
\begin{equation}
  \label{eq:stirap}
  \Omega_j \Delta\tau \gg
  10 \quad (j=B,R)
\end{equation}
is met, where $\Delta\tau$ is the time for
which the pulses overlap. Thus, for short
pulses, large amplitudes are required.  However, for a Rydberg gate, 
the blockade condition, Eq.~\eqref{eq:blockade}, also needs to be fulfilled, limiting 
the maximum Rabi frequency.
Therefore STIRAP can only employ comparatively long pulses for the center $2\pi$
Rabi flip on the right atom.

\begin{figure}[tb] 
  \begin{center}
    \includegraphics{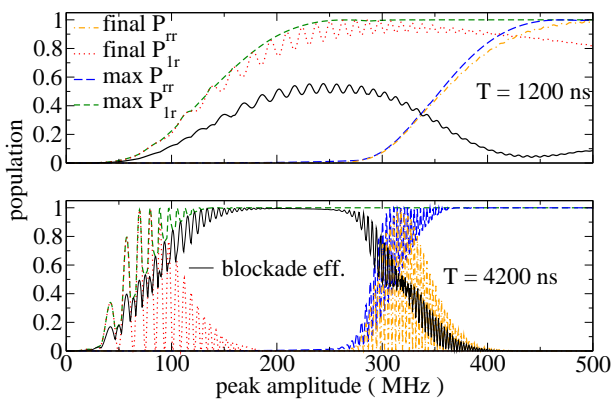}  
  \end{center}
  \caption{(Color online)
    Breakdown of the Rydberg blockade for STIRAP: Only
    long gate durations allow for amplitudes that are sufficiently
    large to ensure adiabaticity in STIRAP while being small enough not
    to break the Rydberg blockade (lower panel).
  }
  \label{fig:blockade_stirap}
\end{figure}
In order to quantify violation of the blockade condition, 
we define the 'blockade efficiency', $B$, to be 
\begin{equation}
  B = \max(P_{1r}) - \frac{1}{2}P_{1r}(T) - 
\left(\max(P_{rr}) - \frac{1}{2}P_{rr}(T)\right)\,,
\end{equation}
where $T$ is the total time of the pulse sequence and $P_{1r}$ and $P_{rr}$ are
the population in $\Ket{1r}$ and $\Ket{rr}$, respectively. $B$ takes values
between zero and one, with one corresponding to a perfect blockade.
Both maximum and final-time populations appear in $B$ because, in order
to have full Rabi cycling, the Rydberg level must be fully populated
(giving a maximum population of one) 
and then fully depopulated (giving a final population of zero), i.e., 
considering only the maximum population does not allow for
distinguishing between $\pi$ and $2\pi$ pulses. We only obtain $B=1$ when 
the population completes a $2\pi$ cycle through $\Ket{1r}$ whenever the system
begins in $\Ket{10}$ but never reaches $\Ket{rr}$ whenever the system begins in
$\Ket{00}$.
The blockade condition, Eq.~\eqref{eq:blockade}, depends on the peak
amplitude of the pulses whereas the 
adiabaticity condition, Eq.~\eqref{eq:stirap},
depends on the pulses' complete Rabi angle. For short pulses
the Rabi angle will not be  sufficiently large 
to satisfy the adiabaticity condition without requiring a peak
amplitude so high that it will break the blockade. This is illustrated in 
Fig.~\ref{fig:blockade_stirap} (top), where 
for small amplitudes both the maximum and final $\Ket{1r}$
populations rise together: the Rabi angle is less than $\pi$ (dashed green and
dotted red lines). Then, as the final 
$\Ket{1r}$ population begins to fall 
such that the adiabaticity 
condition of STIRAP is better fulfilled, the blockade is broken, causing the
drop in the blockade  efficiency, cf. solid black line, concurrent with a
rise in both the maximum and final $\Ket{rr}$ populations (long-dashed blue
and dot-dashed orange lines).
In Fig.~\ref{fig:blockade_stirap} (bottom), the maximum 
and final $\Ket{1r}$ populations rise together
(green dashed and dotted red lines), but 
$\Ket{1r}$ is now fully depopulated, thus achieving full Rabi
cycling, before breaking the 
blockade. This corresponds to the area where $B \approx 1$ seen in the
graph. We do not see a rise in the maximum and final $\Ket{rr}$ population
until high peak amplitudes (long dashed blue and dot-dashed orange lines).

\begin{figure}[tb] 
  \begin{center}
    \includegraphics{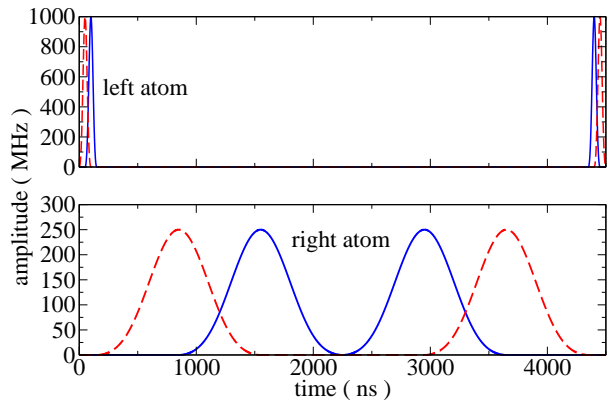}  
  \end{center}
  \caption{A sequence of STIRAP pulse pairs to implement the 
    Rydberg CPHASE gate: While the pulses acting on the left atom
    can be made very short (limited effectively by the power of the driving
    laser), the pulses acting on the right atom need to be sufficiently long to
    avoid breaking the Rydberg blockade.
  }
  \label{fig:Stirap}
\end{figure}
A corresponding  sequence of STIRAP pulse pairs, using short pulses on the left
atom and long pulses on the right atom,  is shown in Fig.~\ref{fig:Stirap}.
In principle, the pulses on the left atom can be made arbitrarily short,
at the expense of extremely large field amplitudes. Taking into account
realistic restrictions on the available laser power, the gate time will
generally become prohibitively large.

\begin{figure}[tb] 
  \begin{center}
    \includegraphics{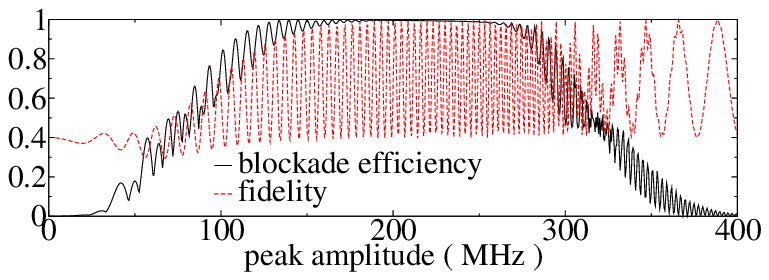}  
  \end{center}
  \caption{(Color online)
    Gate fidelity for the Rydberg gate using STIRAP pulse pairs: even
    for amplitudes for which the blockade condition,
    Eq.~\eqref{eq:blockade}, is fulfilled, the gate fidelity may be
    low due to improper phase alignment. 
  }
  \label{fig:fidelity_stirap}
\end{figure}
Even in the regime where both the adiabaticity condition and the
blockade condition are fulfilled,
the fidelity oscillates rapidly as a function 
of the peak amplitude, as seen in Fig.~\ref{fig:fidelity_stirap}. 
These oscillations are due to phase errors induced by small population of the
intermediate state $\ket{ri}$ during
the central pulses in the blockade regime.
They may be compensated using the techniques proposed in 
Ref.~\cite{beterov2013quantum}, though this does
not address the fundamental issue of large gate time.

\subsection{Mixed scheme: STIRAP-$\pi$-pulses and simultaneous
  $2\pi$-pulses} 
\label{subsec:mixed}

\begin{figure}[tb] 
  \begin{center}
    \includegraphics{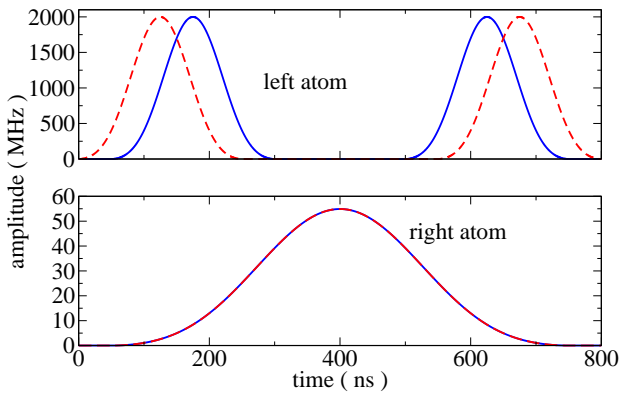}  
  \end{center}
  \caption{(Color online)
    Mixed scheme: STIRAP pulse pairs for robust population
    transfer on the left atom, and simultaneous pulses for the $2\pi$
    rotation of the right atom.
  }
  \label{fig:mixed}
\end{figure}
The primary drawbacks of the simultaneous pulses are the
unwanted population in the intermediate level for the pulses acting on
the left atom and a relatively large sensitivity of 
the pulses to variations in pulse area.
On the other hand, the primary drawback
of STIRAP is the breakdown of the Rydberg blockade, which 
results in employing an extremely long  pulse acting on the right
atom. This issue, however, is  not present when using STIRAP for the
pulses acting on the left atom. We therefore investigate a mixed
scheme, consisting of STIRAP pulses to drive the $\pi$ rotations on
the left atom and simultaneous pulses to drive the $2\pi$ rotation on
the right atom, cf. Fig.~\ref{fig:mixed}. By doing so we 
use each method where it is most effective.
Furthermore, the pulses on the left and right atom can be overlapped 
without any appreciable loss in fidelity. This is because the 
pulses acting on the right atom only drive significant population
transfer during the central third of the pulses. As long as the left
atom is populated by the time the amplitude of the pulses acting on
the right atom becomes significant, the blockade is still effective. The two
STIRAP pulses acting on the left atom, that bookend the central pulses
acting on the right atom, are moved in towards the center. 
In fact the pulses can be compressed
quite far: By overlapping the STIRAP pulses 
with the central pulses for 250$\,$ns, cf. Fig.~\ref{fig:mixed}, 
the gate duration can be
reduced from 1300$\,$ns to 800$\,$ns. The gate duration in the mixed scheme is
limited by the laser power available for driving the left atom.

\subsection{Robustness}
\label{subsec:robust}

For all three variants of pulse sequences, the gate fidelity in an
actual experiment will be compromised by noise and experimental
inaccuracies. We consider in the following three main sources of
errors: inaccuracies in timing between the pulses acting on the left and right
qubit, inaccuracies in pulse
amplitudes, and fluctuations of the Rydberg level due to, e.g., the
presence of DC electric fields~\cite{MuellerKochSpIssue11}. The latter
results in a non-zero two-photon detuning.  
To analyze the robustness with respect to all of these fluctuations, we
determine the expectation value of the gate fidelity under the assumption that
the timing offset, the transition dipole, and the two-photon detuning
differ from the optimal values by $\Delta_{\text{time}}$,
$\Delta_{\Omega}$, and $\Delta_{\text{ryd}}$ drawn from a Gaussian
distribution centered at 0 of width $\sigma_{\text{time}}$, $\sigma_{\Omega}$, and
$\sigma_{\text{ryd}}$, respectively. For the pulse amplitudes,
the variation is given in percent of the original amplitudes. The expectation
value of the gate fidelity is given by
\begin{equation}
  \label{eq:meanF}
  \tilde F(\sigma_x)
  = \int  \frac{1}{\sqrt{2 \pi \sigma_x^2}}
          e^{-\frac{\Delta x}{2 \sigma_x^2}}
          F(\Delta x) \dd x, 
\end{equation}
with $\sigma_{x} = \sigma_{\text{time}}, \sigma_{\Omega}, \sigma_{\text{ryd}}$,
and $\Delta x = \Delta_{\text{time}}, \Delta_{\Omega}, \Delta_{\text{ryd}}$, and
$F$ given by Eq~(\ref{eq:gate_fid}).
Sampling over 1000 variations of each parameter allows to evaluate the
integral in Eq.~(\ref{eq:meanF}) numerically.

\begin{figure}[tb] 
  \centering
  \includegraphics{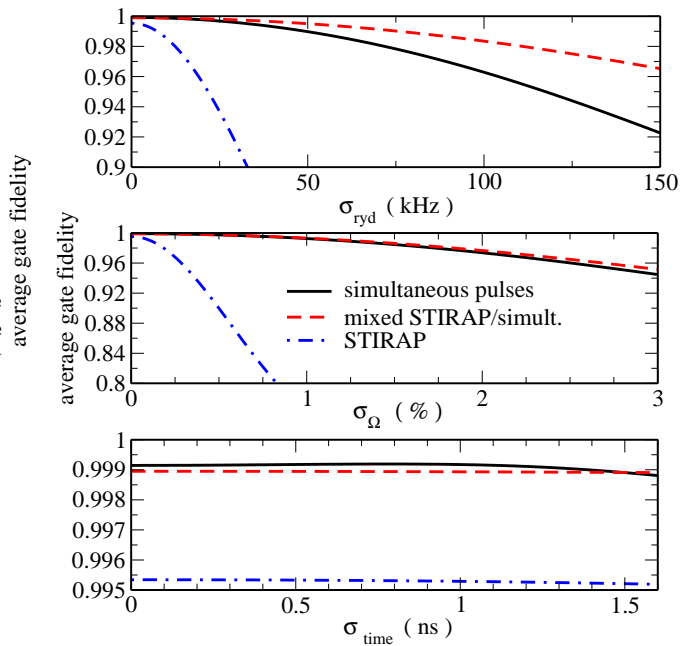}
  \caption{Robustness of the Rydberg gate with respect
    to Rydberg level fluctuations (top), amplitude fluctuations
    (middle), and fluctuations in the relative timing between
    pulses acting on the left and right atom. All fluctuations are 
    drawn from a Gaussian distribution of width $\sigma_{\text{ryd}}$,
    $\sigma_{\Omega}$, and $\sigma_{\text{time}}$, respectively.}
  \label{fig:pulserobust}
\end{figure}
Figure~\ref{fig:pulserobust} shows the resulting expectation value of the 
gate fidelity vs.\ standard 
deviation of the fluctuations in pulse timings, pulse amplitudes, 
and energy of the Rydberg level. 
The gate is found to be very robust with respect to pulse timings 
and fairly robust with respect to amplitudes: only errors of more 
than a few nanoseconds in timing and several per cent in amplitude 
reduce the gate fidelity appreciably.
A larger sensitivity is found with respect to the position of the Rydberg level:
Fluctuations on the order of 1\% of the interaction energy $u$ reduce
the gate fidelity to around 0.5 even for the most robust scheme, and even those
on the order of 0.1\% of $u$ reduce the fidelity appreciably, cf. top
panel of Fig.~\ref{fig:pulserobust}. This is not surprising, since a
'wrong' energy of the Rydberg level leads to a non-zero two-photon
detuning, $\Delta_2$, and thus affects both the population
transfer for the left atom and the non-local phase accumulated during
the pulse acting on the right atom. This additional phase is by assumption
unknown and thus cannot be accounted for. Depending on the choice of the
Rydberg level, the fluctuations of the level energy may be
suppressed down to 100$\,$kHz or less~\cite{Saffman_pc}.
Gate fidelities of about 0.98 are then
within reach, cf.\ the upper panel of Fig.~\ref{fig:pulserobust}. 

Though all the schemes behave similarly to variations in timing, there are
significant differences in each scheme's robustness to fluctuations in
pulse amplitude and Rydberg level energy. For inaccuracies in pulse 
amplitude, cf.\ Fig.~\ref{fig:pulserobust} 
(middle), the fidelity achieved with STIRAP pulses (dot-dashed blue
line) is far more  
susceptible to small variations than both other schemes. This is due to the 
additional phase accumulated for STIRAP during the central  pulse
acting on the right atom, caused by undesired population entering 
$\Ket{ri}$, cf. section~\ref{subsec:stirap}. The mixed scheme (dashed red 
line) performs slightly better than the simultaneous scheme (solid black line) 
in this respect, as the robust STIRAP pulses acting on the left atom
can achieve efficient  population transfer at a wide variety of amplitudes. 
With respect to fluctuations in the energy of the Rydberg level, 
in Fig~\ref{fig:pulserobust} (top) the longer a given scheme populates 
$\Ket{r0}$, the less robust that scheme is. When the population is in the 
detuned $\Ket{r0}$ state, it accumulates  an undesired phase, and this, 
not the loss in population transfer efficiency, is the primary reason for the 
drop in fidelity. The longer a scheme remains in  $\Ket{r0}$, the longer it 
takes to accumulate this additional phase. The mixed scheme, which overlaps 
the pulses acting on the left and right atom and thus populates  
$\Ket{r0}$ for the shortest time possible is the most robust to
fluctuations in the Rydberg level energy. This is followed by the
simultaneous scheme, which fully  populates $\Ket{r0}$ for 700~ns, and 
finally the STIRAP scheme, which fully populates $\Ket{r0}$ for 4200~ns.
Counter-intuitively, then, the schemes actually are less robust with
respect to variations in Rydberg level energy the longer they become. 

\section{Optimal control}
\label{sec:oct}

The use of optimal control theory (OCT) allows to obtain non-analytic pulses
that are not bound by conditions of adiabaticity, and can realize gate
times at the quantum speed
limit~\cite{GoerzJPB11,MuellerKochPRA11,MuellerKochSpIssue11}. 
Here, we extend the application of optimal control to increase the
robustness of the pulses with respect to fluctuations in amplitude and the
energy of the Rydberg level due to external fields. This is achieved by
requiring the gate fidelity, Eq.~(\ref{eq:gate_fid}), to be close
to one not only for the ideal
Hamiltonian $\Op{H}_0$, Eq.~(\ref{eq:2q_ham}), but also for an ensemble
of perturbed Hamiltonians $\{\Op{H}_i\}$, $i=[1,N-1]$ that sample the relevant
parameter space of variations.
Unlike in the analytical pulse schemes, the optimized control pulses will not
consist of sub-pulses, but will be completely overlapping. Therefore, an
analysis of the robustness with respect to pulse timing is not meaningful in
this context.
Since population in the intermediate state
$\Ket{i}$ should be avoided, we perform the optimization in Liouville space, and
include spontaneous emission for the intermediate level with a lifetime of
150$\,$ns~\cite{CampaniLANC1978,OrtizJQSRT1981}. The optimization functional to be
minimized reads
\begin{widetext}
\begin{equation}
  J = 1 - \sum_{n=0}^{N-1} \sum_{i=1}^{16}
    \mathfrak{Re}\left\{\trace\left[
      \Op O\Op\rho_{i}(0)\Op O^{\dagger}\Op\rho_{i,n}\left(T\right)
    \right]\right\}
    - \sum_{j=1}^{4} \lambda_j \int_0^T
      \frac{\left(\Omega_{j}(t)-\Omega_{j, \text{ref}}(t)\right)^2}
      {S(t)}
       \dd t\,,
  \label{eq:functional}
\end{equation}
\end{widetext}
where $\Op{O}$ is the CPHASE gate, up to a trivial global phase due to 
the natural time evolution of the $\Ket{1}$ state; the set of $\Op{\rho}_i$
matrices are the canonical basis elements of the two-qubit Liouville space,
$\{\Ket{i}\!\Bra{j} \} \; \forall i,j \in \{ 00, 01, 10, 11\}$; $\lambda_j$ are
arbitrary positive scaling parameters; the $\Omega_j(t)$ are the four
controls, i.e.\ the fields of the red and blue lasers for the left and
right atom, respectively; 
$S(t)$ is a shape function that ensures a smooth switch-on and switch-off of
the pulses; and the $\Omega_{j, \text{ref}}$ are a set of reference fields.
The gate duration $T$ is fixed for the optimization, but can be systematically
varied in order to determine the quantum speed limit.
For numerical efficiency, the full basis of 16 states can be replaced by just
two density matrices specifically tailored to the optimization problem
\cite{Goerz3States}.
The time dependent states $\Op{\rho}_{i,n}(t)$ are determined by the equation of
motion,
\begin{equation}
  \label{eq:LvN}
  \frac{\partial }{\partial t}\Op{\rho}_{i,n}(t)
  = -i[\Op H_n(t),\Op{\rho}_{i,n}(t)] + \mathcal{L}_D(\Op{\rho}_{i,n}(t))\,
\end{equation}
with $\Op{\rho}_{i,n}(t=0) = \Op{\rho}_{i}(0)$, 
and $\mathcal{L}_D$ the dissipator describing the spontaneous decay
from the intermediate level,
\begin{equation}
  \label{eq:dissipator}
  \mathcal L_D(\Op\rho) = \frac{1}{\tau} \sum_{i=1,2} \left(
    \Op{A}_i \Op{\rho} \Op{A}_i^\dagger
    - \frac{1}{2} \left\{\Op{A}_i^\dagger \Op{A}_i, \Op{\rho} \right\}
    \right)\,,
\end{equation}
with $\Op{A}_1 = \Ket{0}\!\Bra{i} \otimes \unity$, 
$\Op{A}_2 = \unity \otimes \Ket{0}\!\Bra{i}$, and $\tau$ the lifetime.

We use the linear version of Krotov's
method~\cite{PalaoPRA03,ReichKochJCP12} to iteratively  
minimize Eq.~(\ref{eq:functional}).
If the fields from the previous iteration are used as the reference fields
$\Omega_{j, \text{ref}}(t)$, the update equation for each control becomes
\begin{widetext}
\begin{equation}
\label{eq:update}
     \Delta\Omega_j(t) =
     \frac{S(t)}{\lambda_j} \sum_{n=0}^{N-1} \sum_{i=1}^{16} \Im\left\{
     \trace\left(
       -i\,
       \Op\sigma_{i,n}^{\old}(t)
       \left[ 
        \frac{\partial \Op{H}_n}{\partial \Omega_j},
        \Op{\rho}_{i,n}^{\new}(t)
       \right]
     \right)\right\}\,,
\end{equation}
\end{widetext}
with the $\Op\sigma_{i,n}^{\old}(t)$ being a set of co-states backwards
propagated with the pulse from the previous iteration,
\begin{equation}
  \label{eq:backward_prop}
  \frac{\dd \Op{\sigma}_{i,n}(t)}{\dd t}
  = -i[\Op H_n(t),\Op{\sigma}_{i,n}(t)] - \mathcal{L}_D(\Op{\sigma}_{i,n}(t))\,
\end{equation}
and the 'initial' condition
\begin{equation}
  \Op{\sigma}_{i,n}(t=T) = \Op{O} \Op{\rho}_{i}(0) \Op{O}^\dagger\,.
\end{equation}
The states $\rho_{i,n}^{\new}(t)$ are forward propagated using the pulse of the
current iteration, according to Eq.~(\ref{eq:LvN}).
In the case of the rotating wave approximation where the $\Omega_j(t)$ are
complex, Eq.~(\ref{eq:update}) is valid for both the real and the imaginary part
of the pulse.

\begin{figure}[tb] 
  \centering
  \includegraphics{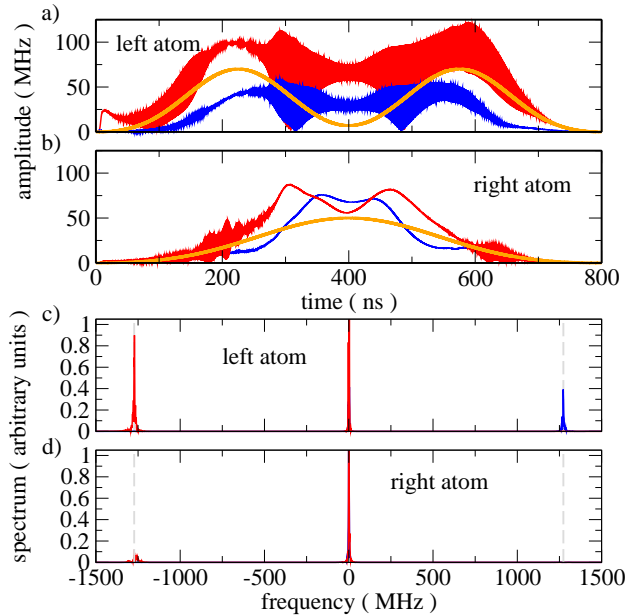}
  \caption{Amplitudes and spectra of pulses optimized with respect to variations
  in both two-photon detuning and pulse amplitude, for a gate duration of
  $T=800$~ns. The central peaks in the
  spectra are truncated to emphasize the side-peaks.
  In panel (c),
  the amplitudes reach a value of 2.0 (red) and 0.8 (blue). In panel (d), the
  peaks reach 2.5 (red) and 3.0 (blue). Pulses and
  spectra are shown in the two-color rotating frame. The central frequency of
  zero corresponds to a laser frequency of the blue pulse that is detuned by
  $\Delta_1$ with respect to the $\Ket{0} \rightarrow \Ket{i}$ transition. For
  the red pulse, it indicates the frequency for which there is a two-photon
  resonance with the $\Ket{0} \rightarrow \Ket{r}$ transition.
  The frequencies matching $\pm \Delta_1$ are indicated by vertical dashed gray
  lines.
  }
  \label{fig:octpulses}
\end{figure}
In order to optimize for robustness with respect to both amplitude fluctuations
and fluctuations of the Rydberg level, we choose an ensemble of $N=24$
Hamiltonians, evenly sampling the values of $\Delta_{ryd}$ between $\pm
300$~kHz and variations of the dipole coupling strength between $\pm 5\%$.
The resulting pulses and their spectra are shown in Fig.~\ref{fig:octpulses}.
The guess pulses from which the optimization started are indicated in orange;
they are inspired by the analytic scheme of the previous section,
consisting of two $\pi$ pulses on the left atom and simultaneously one $2\pi$
pulse on the right atom. The gate duration was set to $T=800$~ns, matching the
shortest gate duration obtained for the analytic schemes in the previous
section.
The choice of the guess pulse is arbitrary in principle,
but has significant impact on the convergence speed and the characteristics of
the optimized pulse. Indeed, the optimized pulse shapes still roughly follow
the shapes of the guess pulses. However, especially for the left atom, there are
fast oscillations present in
the optimized pulse shapes which correspond to a second laser
frequency. As can be seen from the spectra shown in
Fig.~\ref{fig:octpulses}(c), this second  
frequency is at $+\Delta_1$ for the blue pulse and at $-\Delta_1$ for the red
pulse. In sum, these frequencies still make the pulses two-photon resonant with
the $\Ket{0} \rightarrow \Ket{r}$ transition, providing a second excitation pathway whose
interference with the primary pathway can be exploited as a control mechanism.
The blue side peak is smaller simply due to the smaller amplitude of that laser.
In the spectra of the pulses acting on the right atom,
cf.\ Fig.~\ref{fig:octpulses}(d), 
the second frequency is mostly absent, except for the very beginning and end of
the red pulse.
\begin{figure}[tb] 
  \centering
  \includegraphics{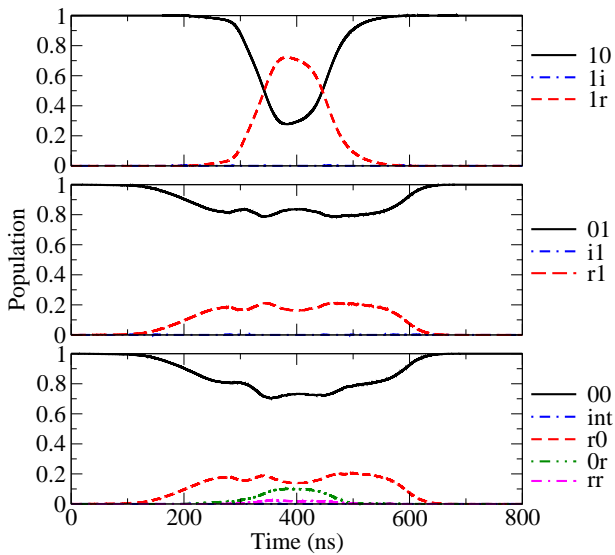}
  \caption{Dynamics under the pulses optimized with respect to fluctuations
  in both the Rydberg level and pulse amplitudes, as shown in
  Fig.~\ref{fig:octpulses}.
  The intermediate population in the bottom panel (''int'') is integrated over
  the states $\Ket{0i}$, $\Ket{i0}$, $\Ket{ii}$, $\Ket{ir}$, and $\Ket{ri}$.
  The shown dynamics implement the desired CPHASE gate up to
  a gate error of $1.04\cdot10^{-4}$.}
  \label{fig:octdyn}
\end{figure}
The population induced by the optimized pulses with the ideal Hamiltonian
$\Op{H}_0$ is shown in Fig.~\ref{fig:octdyn}. Even though the optimized pulses
have frequency components that are resonant with the $\Ket{0} \rightarrow
\Ket{i}$ transition, the intermediate level is never significantly populated, due to
destructive interference. Suppression of the intermediate state population may
be aided by the STIRAP-like feature of the optimized pulse shape, in 
Fig.~\ref{fig:octpulses}(a) and (b), where the red laser
(counter-intuitively) precedes the blue laser in the initial depopulation of the
$\Ket{0}$ level of the left atom, and follows it in the final repopulation.
Furthermore, the population of the $\Ket{01}$ stays remarkable constant, despite
the rather large amplitudes of the laser fields in
Fig.~\ref{fig:octpulses}(a). Again, this is due
to the interfering multiple pathways. In contrast, the dynamics of the
$\Ket{10}$ state is much more straightforward, correspondent to the absence of
the second laser frequency, and consists effectively of a single $2 \pi$ pulse,
although not with full population transfer.
The Rydberg blockade is almost fully maintained, cf.\ the lack of
population in the $\Ket{rr}$ state in the bottom panel of Fig.~\ref{fig:octdyn}.
Also, the right atom in the time evolution of the $\Ket{00}$ state is almost
unaffected by the pulse on the right atom, resulting in very similar
population dynamics for the $\Ket{00}$ and $\Ket{01}$ states.

\begin{figure}[tb] 
  \centering
  \includegraphics{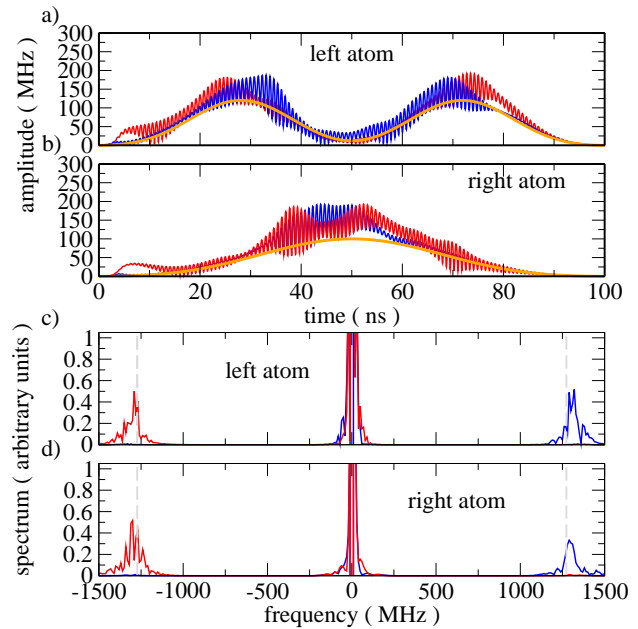}
  \caption{Amplitudes and spectra of pulses optimized with respect to variations
  in both two-photon detuning and pulse amplitude, for a gate duration of
  $T=100~$ns. The spectra are drawn on the same scale as in
  Fig.~\ref{fig:octpulses}, with the central peaks in panel (c) reaching 4.5
  (blue) and 3.0 (red), and 4.5 for both pulses in panel (d).
  }
  \label{fig:octpulses100}
\end{figure}
Optimal control also holds the promise of finding pulses approaching the quantum
speed limit. We can find solutions with gate durations far below $T=800\,$ns
required for the analytic schemes, although very short pulses may require
unfeasibly large pulse amplitudes.
The pulses and spectra resulting from an
optimization for $T=100\,$ns are shown in Fig.~\ref{fig:octpulses100}. The
pulse are optimized for robustness, using the same ensemble of Hamiltonians
as for the $T=800\,$ns pulses. The pulse shapes again follow the features of the
guess pulse, and are only slightly more complex than those for 800$\,$ns in
Fig.~\ref{fig:octpulses}.
The spectra in Fig.~\ref{fig:octpulses100} (c) and (d)  reveal that
a similar mechanism as for $T=800\,$ns is used to produce the gate, through the
presence of additional frequencies at $\pm \Delta_1$. The most significant
difference to Fig.~\ref{fig:octpulses} is that now the additional frequencies
are present for both the left and the right atom throughout the entire gate
duration. 
The peaks in the spectrum are broadened due to the shorter time window. Also, the
pulse amplitudes are now significantly higher. Generally, the optimization
becomes harder for shorter pulse durations, which is why the available control
mechanism must now be used more efficiently, thus causing the presence of the
second laser frequency throughout all pulses.

\begin{figure}[tb] 
  \centering
  \includegraphics{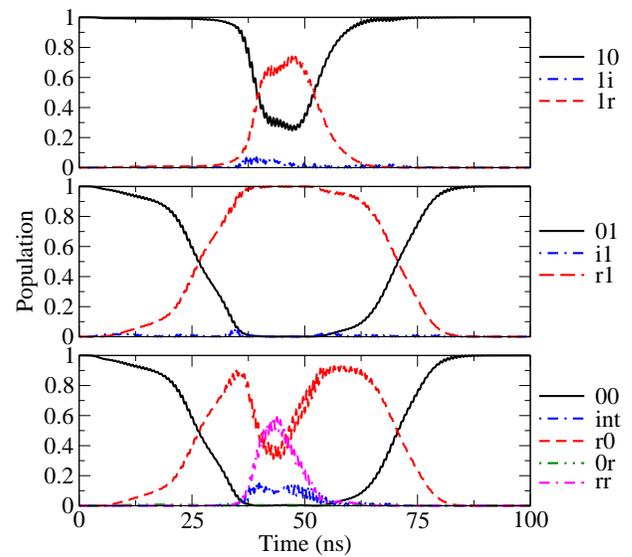}
  \caption{Dynamics under the optimized pulses shown in
  Fig.~\ref{fig:octpulses100}. The gate error is $1.92\cdot10^{-4}$.}
  \label{fig:octdyn100}
\end{figure}
The population dynamics, shown in Fig.~\ref{fig:octdyn100}, reflect the
increase in the laser amplitudes through some significant differences compared to the
dynamics shown in Fig.~\ref{fig:octdyn}. Most importantly, the Rydberg blockade
is now broken, resulting in a significant population of the $\Ket{rr}$ state,
cf.\ the purple curve in the bottom panel. This nicely illustrates the power of
OCT; while the analytic schemes rely on maintaining the blockade regime, the
optimization has no such restrictions, and will exploit any pathways available
in the time evolution generated by the two-qubit Hamiltonian.
There is some minor population in the intermediate
states during the propagation of the $\Ket{00}$ state, cf.\ the blue line in the
bottom panel of Fig.~\ref{fig:octdyn100}. However, since the dynamics result
from an optimization that took into account the spontaneous decay from the
intermediate level explicitly, we are guaranteed that the population in this
level is below a threshold that will affect the gate fidelity.

\begin{figure}[tb] 
  \centering
  \includegraphics{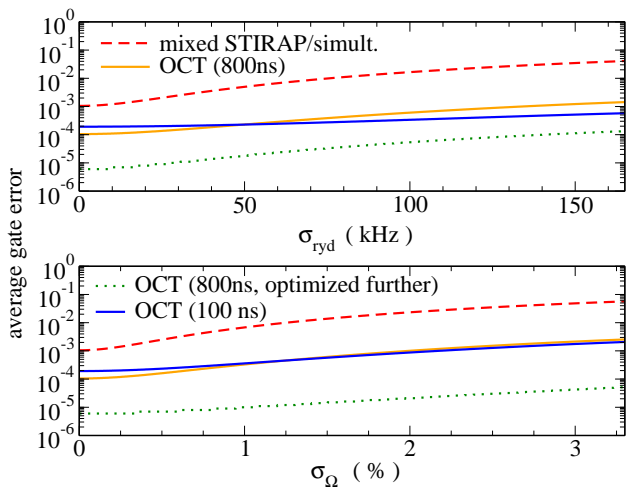}
  \caption{Expectation value of the gate error in the presence of fluctuations
    in the $\Ket{rr}$ state due to DC electric fields (top), and pulse amplitude
    fluctuations (bottom). The red dashed curve shows the most robust analytical
    pulse, cf.\ the red dashed curve in Fig.~\ref{fig:pulserobust}. The solid
    yellow and blue lines are for the optimized pulses shown in
    Figs.~\ref{fig:octpulses},~\ref{fig:octpulses100}, respectively. The dotted
    green line is for a further optimized pulse at $T=800$~ns, without
    any consideration of limits on the pulse amplitude or complexity.
    Note that both panels show the robustness for same set of pulses, i.e.\ the
    pulses were optimized with respect to \emph{both} variations in the
    two-photon detuning and the pulse amplitude.
    }
  \label{fig:robustoct}
\end{figure}
In Fig.~\ref{fig:robustoct}, we compare the effect of fluctuations on
the gate fidelity for the pulses obtained with OCT, cf.\
Figs.~\ref{fig:octpulses} and~\ref{fig:octpulses100}, to that for the most
robust gates achieved with the analytic schemes, i.e.\ the 
mixed scheme employing STIRAP for the pulses on the left atom, and simultaneous
pulses for the right atom, cf.\ Fig.~\ref{fig:pulserobust}.
The optimized pulses are significantly more robust with respect to both sources
of error by at least an order of magnitude, with the gate fidelity staying above
99.9\% even for large variations, whereas for the analytic pulses, it drops
below 97\% for fluctuations of the Rydberg level (top panel) and 95\% for
amplitude fluctuations (bottom panel).
Note that in contrast to the analytic mixed scheme, the optimized pulses do not
require unfeasibly large pulse amplitudes. 
In contrast, the scheme using only simultaneous
pulses but more realistic pulse amplitudes would be even more sensitive --
particularly to fluctuations of the Rydberg level (cf.\ the drop to 92\% gate
fidelity in the top panel of Fig.~\ref{fig:pulserobust}). The price for
this additional robustness offered by the numerically optimized pulses is
a slightly more complex pulse shape and the
presence of a second frequency.

It is important to note that the solutions provided by OCT are not unique; the
pulses obtained depend on the guess pulses, the exact choice of optimization
functional, and on arbitrary scaling parameters such as the $\lambda_j$ in
Eq.~(\ref{eq:functional}). By tuning these parameters carefully, the
optimization may be steered towards desired pulse features.
It is also possible to add additional constraints to the
optimization functional in order to preselect optimization
pathways~\cite{JosePRA13}. For example, the $\Ket{rr}$ state could
be defined as a forbidden subspace in order to enforce the blockade regime, if
so desired. One could also include spectral
constraints to impose a prespecified pulse bandwidth or suppress
undesired frequency components~\cite{JosePRA13,ReichKochJMO13}. 
Optimizing to extremely high fidelities often leads to very large
pulse amplitudes or complex pulse features that are undesirable from an
experimental point of view. Thus, it is usually best to stop the optimization as
soon as the reached fidelities are ``good enough'', as was done for the
optimized pulses shown as solid blue and yellow lines in
Fig.~\ref{fig:robustoct}.
In principle, however, pulses of much higher fidelity
and robustness than those shown here can be found.  This is illustrated by the
dotted green line in Fig.~\ref{fig:robustoct}, which shows the result of
a further optimization of the pulse for $T=800$~ns. 
While these pulses achieve a 
gate fidelity well above that required for fault tolerant
quantum computation~\cite{gottesman2013overhead,reichardt2009error},
the resulting highly optimized pulses
have unfeasibly large pulse amplitudes of 1100~MHz and 330~MHz for the blue
and red laser, respectively.

\section{Summary and conclusions}
\label{sec:concl}

We have studied  high-fidelity controlled
phasegates based on the Rydberg blockade and investigated their robustness 
with respect to noise due to stray fields causing fluctuations of the
Rydberg level as well as experimental inaccuracies in pulse timings and
amplitudes. 
When single site addressability is available, the gate can be
completed by a $2\pi$-pulse on the right atom, preceded and followed
by a $\pi$-pulse on the left atom.
For practical reasons, the excitation to the Rydberg level uses a
two-photon transition, i.e., each of the three pulses is replaced by
a pair of pulses with different frequencies. The pulse pairs can be
chosen to occur simultaneously or time-delayed, the latter mimicking a STIRAP
sequence. For simultaneous pulse pairs, the Rabi frequency of the red and
the blue laser must be identical to achieve population
inversion~\cite{ShoreBook11}. This is not required for STIRAP. 
The shortest possible gate duration with analytical pulse shapes is
found for a combination of STIRAP pulses acting on the left atom and
simultaneous pulses acting on the right atom. The gate duration is
limited by the blockade condition which restricts the peak amplitude
of the pulses. The STIRAP pulses must furthermore fulfill the
adiabaticity condition whereas the peak amplitude of the simultaneous
pulses is restricted by the requirement of adiabatic elimination of
the intermediate level. The gate duration can be significantly
shortened by utilizing numerical optimal control to determine the
pulse pairs. In this case,  neither the
blockade condition nor the adiabaticity condition are relevant, and
the gate duration is limited by the strength of the interaction
between two Rydberg atoms. 

For an ideal implementation of the pulse sequences, very high
fidelities beyond the quantum error correction threshold can be
achieved. This is, however, severely compromised when noise and
experimental inaccuracies are taken into account. Gates consisting
of STIRAP pairs for all three pulses are found to be the most
susceptible to noise with amplitude errors of less than 1\% reducing
the fidelity to only 0.8. This surprising result is explained by the
sensitivity of the gate to proper phase alignment: While STIRAP
ensures robust population transfer, additional corrections are
required to compensate undesired phase evolution~\cite{beterov2013quantum}. 
Simultaneous pulses and a combination of
STIRAP and simultaneous pulses are somewhat more robust. However, also
for these pulse sequences, the fidelities are reduced to below 0.95
for realistic noise levels. Of the three noise sources considered,
fluctuations of the Rydberg level due to stray fields are the most
severe, whereas timing inaccuracies of the order of 1$\,$ns play
almost no role. 

In order to identify pulse sequences that are inherently robust to
noise, we have employed optimal control theory and calculated pulses
which guarantee a high gate fidelity as long as the fluctuations of
the Rydberg level and pulse amplitude are confined to a predefined
tolerance window. For realistic noise levels we were able to generate
pulses that yield gate errors well below $10^{-3}$,
with errors  below $10^{-5}$ being reached when no limits are placed on
pulse amplitudes.
Optimized pulse sequences are not only more robust but can also be 
of much shorter duration.
For both short and long gates, the
optimized pulses require only one more frequency, corresponding to the
one-photon detuning, and their temporal
shape is comparatively simple. 
Taking into account the restrictions of feasible pulse amplitudes, we therefore
conclude that 
optimized pulses achieving fault
tolerant gates are experimentally realizable. 

Our optimized pulses may also point the way for the construction of
improved analytical pulse sequences. The additional frequencies
identified by the optimization are utilized to build destructive
interference in the intermediate level that is most severely affected
by spontaneous decay. It thus allows for resonant transitions, 
decreasing the pulse amplitudes and lifting the requirements due to
adiabaticity and adiabatic elimination. 

Lastly, the optimization technique presented here may also be able to address
robustness of the gate with respect to undesired excitation in the vibrational
degree of freedom, which was not considered in this paper. A vibrational
excitation could be modeled as a fluctuation of the $\Ket{0}$ or intermediary
level, which would allow to optimize over an ensemble of different vibrational
states.

\begin{acknowledgments}
  We thank Dave Weiss, Ted Corcovilos, Mark Saffman and Reinhold Bl\"umel 
  for fruitful discussions. 
  This research was supported by the DARPA QuEST program, the Deutscher Akademischer
  Austauschdienst and the National Science Foundation under the Catalzying International Collaborations program (Grant No. OISE-1158954).
   We also thank the Kavli Institute for Theoretical Physics for hospitality and for supporting this 
  research in part by the National Science Foundation Grant No. PHY11-25915.
\end{acknowledgments}


\end{document}